\documentclass[prl,twocolumn,superscriptaddress,showpacs]{revtex4}
\def\beq{\begin{equation}}
\def\eeq{\end{equation}}
\usepackage{amsmath}
\usepackage{graphicx}
\usepackage{epsfig}
\usepackage[usenames]{color}
\newcommand{\bra}[1]{\langle {#1} |}
\newcommand{\ket}[1]{| {#1} \rangle}

\begin{document}
\title{Theory of finite-entanglement scaling at one-dimensional quantum critical points}

\author{Frank Pollmann$^1$, Subroto Mukerjee$^{1,2}$, Ari Turner$^1$ and Joel E.
Moore}

\affiliation{Department of Physics, University of California, Berkeley}
\affiliation{Materials Science Division, Lawrence Berkeley National Laboratory}

\begin{abstract}
Studies of entanglement in many-particle systems suggest that most quantum critical ground states have infinitely more entanglement than non-critical states.  Standard algorithms for one-dimensional many-particle systems construct model states with limited entanglement, which are a worse approximation to quantum critical states than to others.  We give a quantitative theory of previously observed scaling behavior resulting from finite entanglement at quantum criticality: the scaling theory of finite entanglement is only superficially similar to finite-size scaling, and has a different physical origin.  We find that finite-entanglement scaling is governed not by the scaling dimension of an operator but by the ``central charge'' of the critical point, which counts its universal degrees of freedom.  An important ingredient is the recently obtained universal distribution of density-matrix eigenvalues at a critical point\cite{calabrese1}.  The parameter-free theory is checked against numerical scaling at several quantum critical points.
\end{abstract}

\maketitle

A system in its ground state can undergo a second-order (continuous) quantum phase transition as a control parameter is varied through a critical value.  As at a thermal phase transition, the critical point has correlations over long length scales. These correlations are described by properties of the critical point that are ``universal'', i.e., independent of microscopic details.  The entanglement entropy is a measure of the quantum-mechanical nature of correlations and in many cases is also universal~\cite{amico,wilczek,osterloh,vidalent,calabresecardy,refaelmoore,ryutakayanagi,fradkinmoore}.  The entanglement entropy of a pure state of a bipartite system $AB$ is defined as
\beq
S = -{\rm Tr} \rho_A \log \rho_A = -{\rm Tr} \rho_B \log \rho_B,
\eeq
where $\rho_A$ ($\rho_B$) is the reduced density matrix of subsystem $A$($B$).  A nonzero entanglement entropy reflects that in quantum mechanics, a ``complete description'' of a system (i.e., a pure state) does not imply a complete description of subsystems.

We would like to understand a consequence of diverging entanglement at quantum criticality: any approach constructing states with a limited amount of entanglement will show universal, systematic errors in describing quantum critical states.  It was shown previously by Tagliacozzo et al.~\cite{latorre} in a numerical study of two one-dimensional critical models that finite entanglement leads to scaling behavior like that induced by other perturbations of a critical point: it introduces a finite correlation length $\xi \sim \chi^\kappa$, where $\chi$ (defined below) is related to how much entanglement is retained and $\kappa$ is the finite-entanglement scaling exponent.  That work gave convincing evidence for this behavior in the two models studied but did not attempt to explain its origin or develop a theory predicting $\kappa$.  As the retained entanglement $\chi$ increases, $\xi \rightarrow \infty$ and criticality is restored.

The main result of this Letter is a theory for this behavior for conformally invariant critical points in one dimension.  The theory predicts that $\kappa$, unlike other scaling exponents, is determined by the central charge of the critical point.  It leads to a specific formula for this dependence and also explains the observed scaling of entanglement entropy.  The iTEBD algorithm~\cite{vidal}, which has finite-entanglement errors but not finite-size ones, is then used to study a number of critical points.  The numerical results confirm that central charge determines finite-entanglement scaling and show parameter-free agreement with the theoretical predictions.


The best understood quantum critical points are in one spatial dimension, where most translation-invariant systems have critical points with ``conformal invariance'', an infinite symmetry group related to conformal maps of the plane.  The entanglement entropy between two halves of a large \emph{one-dimensional} system close to the critical point (large correlation length $\xi$) is \cite{calabresecardy}:
\begin{equation}
S=\frac{c}{6}\log (\xi/a).
\label{eq:defs1}
\end{equation}
Here $c$ is the ``central charge" of the critical point, a number that counts how many degrees of freedom of the system are critical, and $a$ is a short-distance length scale, such as the lattice spacing in a spin chain.

Our goal is to understand how the increased entanglement of a quantum critical state (Eqn.~(\ref{eq:defs1})) affects classical representation of the state, e.g., in a computer algorithm.  A powerful approach to obtain low-energy states of one-dimensional systems is the density-matrix renormalization group (DMRG) algorithm \cite{white}.  This algorithm and its descendants~\cite{vidal} construct trial wavefunctions that are ``matrix product states'' (MPS) \cite{ostlundrommer}.  Consider a system with $N$ sites, where each site has $d$ orthogonal states.  Any pure state of the system is a superposition of the product basis states $\ket{s_1s_2\dots s_N}=\ket{\{s\}}$ where $1\leq s_i\leq d$.  A MPS for such a system has the form
\begin{equation}
|\psi\rangle=\sum^d_{s_1,\dots, s_{N}=1}\mathrm{Tr}\left[{A^{[1]}_{s_1}\dots A^{[N]}_{s_{N}}} \right]|s_1\rangle\dots|s_{N}\rangle.
\label{eq:mps}
\end{equation}
For each site $i$ there are $d$ matrices $A^{i}_{s_i}$ of a finite dimension $\chi \times \chi$. A product wavefunction describing an unentangled chain of particles/spins is obtained by multiplying together scalar amplitudes for the particle/spin state at each site.  The MPS generates entanglement by using matrices instead of amplitudes for each site, and an increasing amount of information can be stored as the matrix dimension $\chi$ increases.

A convenient representation of this state $\ket{\Psi}$ for entanglement purposes is the Schmidt decomposition.
Splitting the system into two parts at one bond, the Schmidt decomposition is a basis choice that expresses the original wavefunction as a sum of product states of wavefunctions for the two halves of the system:
\begin{equation}
\ket{\Psi}=\sum_{n=1}^{\infty} \lambda_{n} \ket{\Phi_{n A}}\ket{\Phi_{n B}}.
\label{eq:schmidt}
\end{equation}
The $\ket{\Phi_{n A}}$ and $\ket{\Phi_{n B}}$ form orthonormal bases
for the Hilbert spaces of the subsystems  $A$ and $B$ to the left and right of the bond.
The Schmidt decomposition contains more than one term for entangled states such
as an Einstein-Podolsky-Rosen state \cite{einstein}: measurements on subsystem $B$ put the system in a state $\Phi_{n B}$ with probability $\lambda_{n}^2$ and force the left half of the system into the corresponding basis state.  This converts
the system into a \emph{mixed} state with entropy
\begin{equation}
S=-\sum_{n} \lambda_{n}^2 \log \lambda_{n}^2.
\label{eq:defs2}
\end{equation}

We use the recently developed infinite Time-Evolving Block Decimation (iTEBD) algorithm \cite{vidal} to study several local one-dimensional Hamiltonians with translationally invariant ground states.  Exploiting translational invariance, there are only finitely many different matrices $A^i_{s_i}$, which are found by minimising the energy of the wave function in Eqn.~(\ref{eq:mps}).  Because this method always constructs wavefunctions for the infinite system, its errors result from finite entanglement rather than finite size; it was similarly used in~\cite{latorre} for the original numerical study of how finite entanglement affects quantum criticality.  Here our focus is a theory for how the numerical observations differ fundamentally from conventional finite-size scaling, together with additional iTEBD results.  We study cases in which the iTEBD algorithm converges effectively to the matrices of a given dimension that minimize the expectation value of the Hamiltonian \cite{vidal}.  Away from a critical point, the entanglement entropy of the exact wavefunction is finite, and the approximation by matrix product states converges rapidly~\cite{arealaw}.

At the critical point, finite-dimensional matrices cannot approximate the ground state as well; the maximum entropy for $\chi$ Schmidt eigenvalues is $\log \chi$. The numerical eigenvalues for the system at the critical point but with finite $\chi$ actually obey a distribution that does not maximize the entropy.  Instead, there is a physical interpretation of the observed distribution, which underlies the theoretical analysis.  We first note that the observed eigenvalue distribution at finite $\chi$ and at the critical point (correlation length $\xi=\infty$) is similar to the infinite-$\chi$ distribution away from the critical point (at some finite $\xi$ determined by $\chi$).  This is the microscopic origin of the observation \cite{latorre} that physical quantities computed at infinite $\xi$ and finite $\chi$ reproduce those at finite $\xi$ and infinite $\chi$, with an empirical scaling law $\xi \propto \chi^\kappa$ in which the exponent $\kappa$ depends on the specific critical point.  More precisely,
the eigenvalue distribution of the reduced density matrix of a one-dimensional system with a large correlation length $\xi$ was recently studied \cite{calabrese1}: the mean number of eigenvalues larger than a given value $\lambda$ is
\begin{equation}
n(\lambda,b)=I_0(2\sqrt{b(-b-2\log\lambda)}).
\label{eq:CL}
\end{equation}
Here $I_0$ is the zeroth modified Bessel function and $b$ is a parameter that varies with $g$, diverging at the critical point; in fact, summing Eqn.~(\ref{eq:defs2}) shows that $b$ is determined by the entanglement entropy: $b=S/2$. We find that this distribution is a good description of the observed eigenvalue distributions, even at criticality, with finite but large $\chi$.   We now discuss a theory for the finite-entanglement scaling exponent $\kappa$,
which also appears in the scaling of entanglement entropy with $\chi$,
\beq
S=\frac{c\kappa} {6}\log\chi = {c \over 6} \log \xi.
\label{eq:sscal}
\eeq

We wish to compute the effective correlation length $\xi$ that describes the state obtained by minimization of energy at the critical point and finite $\chi$.  Since the true ground state has infinite $\xi$, a state with a finite $\xi$ carries an energy cost proportional to $1/\xi^2$ \cite{supinfo}.  (We assume that the critical point separates two gapped phases, rather than terminating a critical phase.)  However, this energy cost is balanced because the retention of only the first $\chi$ eigenvalues has a more severe effect on a $\xi=\infty$ critical state than a noncritical state.  Hence the energy ($E$) for finite $\chi$, as a function of correlation length $\xi$, should include a second term that computes the energy cost from discarding all but the largest $\chi$ eigenvalues:
\begin{equation}
E(\xi) = E_{\infty}+\frac{A}{\xi^2} + \beta \frac{P_r(b,\chi)}{\xi}.
\label{Eq:enerpert}
\end{equation}
Here $E_{\infty}$ is the ground-state energy of the critical system per unit length and $P_r(b,\chi)=\sum_{n=\chi}^\infty \lambda (b,n)^2$ is the residual probability, making the $b$ dependence explicit. Note that $\beta$ and $A$ are non-universal constants that will drop out of the asymptotic behavior for large $\chi$ \cite{supinfo}.

In the asymptotic large-$\chi$ limit, the discarded eigenvalues can be assumed to form a continuum and be distributed according to Eqn.~(\ref{eq:CL}). The residual probability is now an integral with the values of $\lambda (b,n)$ obtained by inverting Eqn.~(\ref{eq:CL}). For large-$\chi$, we find
\begin{equation}
P_r(b,\chi) = \frac{2be^{-b}\chi}{\log \chi - 2b}e^{\frac{-(\log \chi)^2}{4b}},
\label{Eq:expPr}
\end{equation}
Replacing $\xi$ in favor of $b$ using
$\xi=e^{6S/c}=e^{12b/c}$ and using Eqn.~(\ref{Eq:expPr}) in Eqn.~(\ref{Eq:enerpert}) gives the energy per unit length as a function of $b$ and $\chi$.  To find the ground state at fixed $\chi$, we minimise the energy with respect to $b$ and find that
\begin{equation}
\kappa = \frac{6}{c\left(\sqrt{\frac{12}{c}}+1\right)}\Rightarrow S=\frac{1}{\sqrt{\frac{12}{c}}+1}\log\chi,
\label{Eq:musol}
\end{equation}
with corrections of order $1/\log \chi$.  The relationship between $\kappa$ and $c$ is a central result of this Letter.

We have performed numerical tests of the scaling prediction in Eqn.~(\ref{Eq:musol}) on several critical points. The specific one-dimensional systems we have studied are the quantum Ising model in a transverse field, the $XXZ$ spin chain for spin-1/2, and the spin-1 generalized Heisenberg model. The quantum Ising model whose Hamiltonian is
\begin{equation}\nonumber
H = -\sum_i \left(\sigma_i^x \sigma_{i+1}^x + g \sigma_i^z\right),
\label{Eq:hamising}
\end{equation}
is critical at $g=1$ with central charge $c=1/2$. The $XXZ$ spin chain is described by the Hamiltonian
\begin{equation}\nonumber
H = \sum_i \left(\sigma_i^x \sigma_{i+1}^x + \sigma_i^y \sigma_{i+1}^y + \gamma \sigma_i^z\sigma_{i+1}^z\right)
\label{Eq:hamxxz}
\end{equation}
and is critical in the entire range $\gamma \in [0,1]$ with central charge $c=1$. Critical exponents change continuously with $\gamma$ \cite{baxter}.
The last system we have studied is the $S=1$ Heisenberg chain with biquadratic term,
\begin{equation}\nonumber
H = \sum_i \left[\cos \theta ({\bf S}_i {\bf .} {\bf S}_{i+1}) + \sin \theta ({\bf S}_i {\bf .} {\bf S}_{i+1})^2\right].
\label{Eq:hamspin1}
\end{equation}
This system is known to have two exactly solvable critical points, the SU$(2)_2$ point at $\theta=-\pi/4$ with $c=3/2$ \cite{alcaraz} and the SU$(3)_1$ point at $\theta=\pi/4$ with $c=2$ \cite{itoi}. The entire region $\theta \in [\pi/4, \pi/2)$ is critical with $c=2$ \cite{lauchli}, which is consistent with our results. A detailed study of critical SU($N$) can be found in Ref.~\cite{rachel}.

\begin{figure}
\includegraphics[width=3.5in]{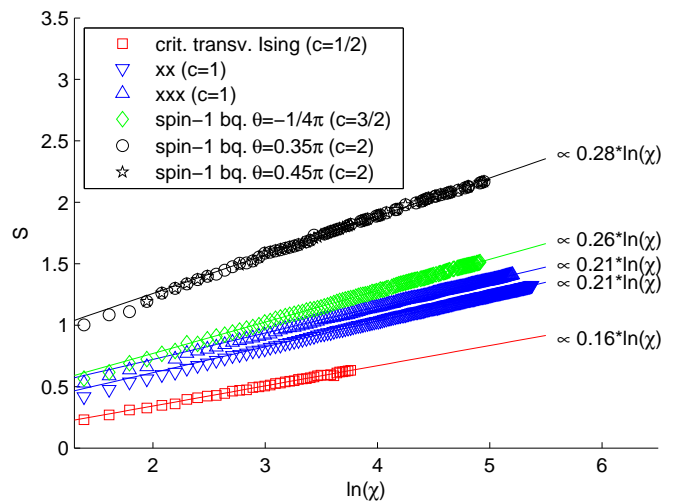}
\caption{Scaling of the entropy with matrix dimension $\chi$: The entropy calculated numerically using the iTEBD algorithm for several critical points in one dimension: the XX and Heisenberg (XXX) models, the transverse Ising model, and the $s=1$ biquadratic Heisenberg model (cf. Methods). It can be seen that in each case $S \propto \log \chi$. Furthermore, the lines for models with the same central charge $c$ have the same slope, although other universal properties differ.}
\label{Fig:ent1}
\end{figure}

\begin{figure}
\includegraphics[width=3.5in]{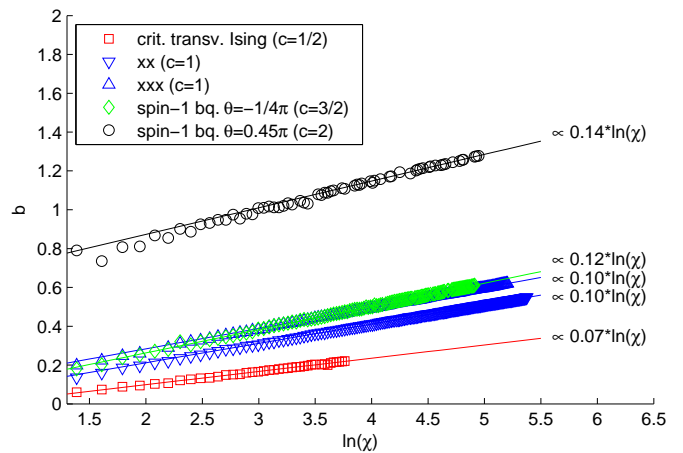}
\caption{Scaling of the parameter $b$: $b=- \log \lambda_{max}$ calculated numerically for different models, where $\lambda_{max}$ is the largest eigenvalue of the Schmidt decomposition. For all models, $b \propto \log \chi$ for sufficiently large matrix dimension $\chi$.  Comparison to Fig.~\ref{Fig:ent1} shows that the relationship $S=2b$ is well satisfied.}
\label{Fig:ent2}
\end{figure}

\begin{figure}
\includegraphics[width=3.5in]{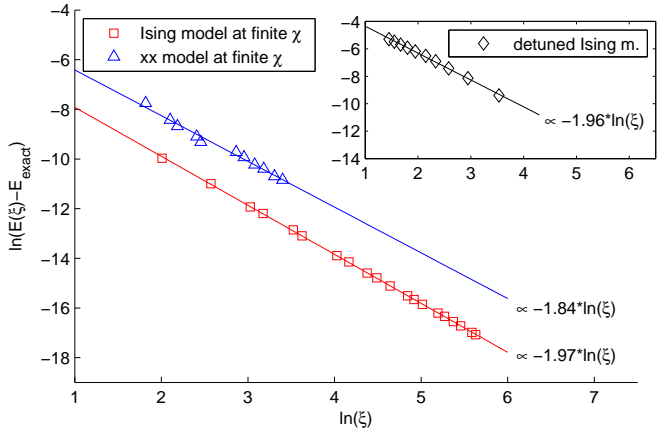}
\caption{Energy vs. correlation length: The energy calculated numerically scales as $1/\xi^2$ with correlation length $\xi$.  The same scaling also appears (inset) when the state in question is obtained as the ground state of a slightly off-critical Hamiltonian.  The correlation length in a matrix product state can be obtained from the transfer operator \cite{wolf}.}
\label{Fig:energyxi}
\end{figure}

The prediction that the scaling of the entropy $S$ depends only on $c$ can be checked directly (Fig.~\ref{Fig:ent1}).  The relationship between the entropy and the largest density-matrix eigenvalue, which is used by Calabrese and Lefevre \cite{calabrese1}, is approximately satisfied (Fig.~\ref{Fig:ent2}).  The model above implies that the energy in the critical Hamiltonian for a state with finite value of $\xi$ should go as $E = E_0 + B/\xi^2$, where $E_0$ is the actual ground-state energy and $B$ a non-universal constant.  This connection between the actual correlation length $\xi$ and the energy was tested (Fig.~\ref{Fig:energyxi}) for ground states on and off criticality.

\begin{figure}
\includegraphics[width=3.5in]{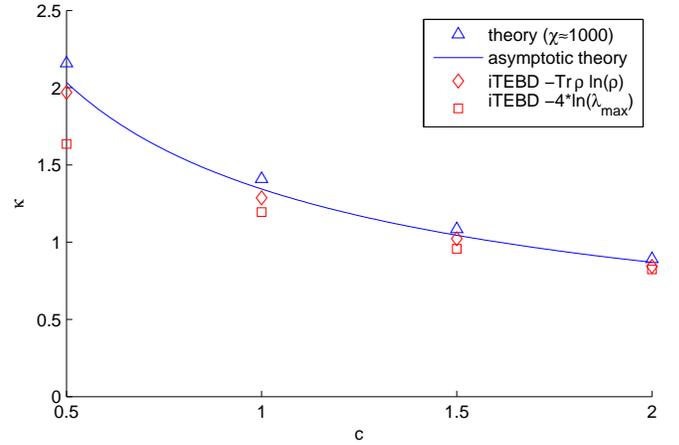}
\caption{$\kappa$ for different values of the central charge: $\kappa$ calculated numerically using different definitions of the entropy (equations (\ref{eq:defs1}) nearly agree (\ref{eq:defs2}), while $S=2b$ gives different values as shown), and according to the parameter-free asymptotic theoretical result in Eq.~(\ref{Eq:musol}). Numerical evaluations of the theoretical model for finite $\chi$ are also shown.
One cause of discrepancies between the values corresponding to different definitions of the entropy is that only finite $\chi$ can be studied; the different definitions of entropy should converge in the asymptotic large-$\chi$ limit.  The numerical values of $\kappa$ at $c=1/2$ and $c=1$ are consistent with those reported in Ref.~\onlinecite{latorre}}
\label{Fig:muvsc}
\end{figure}

Numerical calculations can estimate the entropy scaling using the definition of the entropy in Eqn.~(\ref{eq:defs1}) as well as the relations Eqn.~(\ref{eq:defs2}) and $S=2b$~\cite{S2brefs}, and compare the results to the parameter-free theoretical prediction Eqn.~(\ref{Eq:musol}).
The results are shown in Fig.~\ref{Fig:muvsc}.  The agreement of the numerical values among themselves and with Eqn.~(\ref{Eq:musol}) is at worst about 20\%; even for small $\chi$ the error in the asymptotic theoretical prediction is comparable to the discrepancies between definitions of the entropy.
A stringent test of the nonlinear $c$ dependence of Eqn.~(\ref{Eq:musol}) is that systems of decoupled copies are also found numerically to obey this scaling prediction (not shown).


These results suggest that the effect of finite entanglement in any MPS description near a quantum critical point in one dimension results from universal properties of the quantum critical point, specifically its central charge rather than a scaling dimension as in finite-size scaling.  There are several potential extensions of this approach.  In addition to studying other types of quantum critical points, there are recently developed MPS-type algorithms for time-dependent problems \cite{schollwock} and higher dimensions \cite{peps} that may also show universal scaling of errors near quantum criticality.





\begin{thebibliography}{18}
\bibitem{calabrese1} {P. Calabrese and A. Lefevre, {Phys. Rev. A} {\bf 78}, 032329 (2008).}
\bibitem{amico}{L. Amico, R. Fazio, A. Osterloh, and V. Vedral, {Rev. Mod. Phys.} {\bf 80}, 517 (2008).}
\bibitem{wilczek} {C. Holzhey, F. Larsen, and F. Wilczek, {Nucl. Phys.} {\bf B424}, 443 (1994).}
\bibitem{osterloh} {A. Osterloh, L. Amico, G. Falci, and R. Fazio, {Nature} {\bf 416}, 608 (2002).}
\bibitem{vidalent}{G. Vidal, J. I. Latorre, E. Rico, and A. Kitaev, {Phys. Rev. Lett.} {\bf 90}, 227902 (2003).}
\bibitem{calabresecardy}{P. Calabrese and J. Cardy, {J. Stat. Mech.} {\bf 06}, P06002 (2004).}
\bibitem{refaelmoore} {G. Refael and J. E. Moore, {Phys. Rev. Lett.} {\bf 93}, 260602 (2004).}
\bibitem{ryutakayanagi} {S. Ryu and T. Takayanagi, {Phys. Rev. Lett.} {\bf 96}, 181602 (2006).}
\bibitem{fradkinmoore} {E. Fradkin and J. E. Moore, {Phys. Rev. Lett.} {\bf 97}, 050404 (2006).}
\bibitem{latorre} {L. Tagliacozzo, T. R. de Oliveira, S. Iblisdir, and J. I. Latorre, {Phys. Rev. B} {\bf 78}, 024410 (2008).}
\bibitem{vidal} {G. Vidal, {Phys. Rev. Lett.} {\bf 98}, 070201 (2007).}
\bibitem{white} {S. R. White, {Phys. Rev. Lett.} {\bf 69}, 2863 (1992).}
\bibitem{ostlundrommer} {S. \"Ostlund and S. Rommer, {Phys. Rev. Lett.} {\bf 75}, 3537 (1995).}
\bibitem{einstein} {A. Einstein, B. Podolsky, and N. Rosen, {Phys. Rev.} {\bf 47}, 777 (1935).}
\bibitem{baxter} {R. Baxter, {\em Exactly solved models in statistical mechanics}, Academic, New York (1982).}
\bibitem{alcaraz} {F. C. Alcaraz and M. J. Martins, {J. Phys. A: Math. Gen.} {\bf 21} L381 (1988).}
\bibitem{itoi} {C. Itoi and M.-H. Kato {Phys. Rev. B} {\bf 55} 8295 (1997).}
\bibitem{arealaw} {J. Eisert, M. Cramer, and M. B. Plenio, arXiv:0808.3773, Rev. Mod. Phys. to appear.}
\bibitem{supinfo} {The full derivation of this result for the energy is provided as supplementary information in EPAPS.  The $1/\xi^2$ dependence is a general feature of quantum critical points with conformal invariance.}
\bibitem{wolf} {M.~W. Wolf et al., {Phys. Rev. Lett.} {\bf 97}, 110403 (2006).}
\bibitem{lauchli} {A. L\"auchli, G. Schmid and S. Trebst, {Phys. Rev. B} {\bf 74}, 144426 (2006).}
\bibitem{rachel}{M. F\"uhringer, S. Rachel, R. Thomale, M. Greiter, and P. Schmitteckert, Ann. Phys. (Berlin) {\bf 17}, 922 (2008)}
\bibitem{S2brefs}{J. Eisert and M. Cramer, Phys. Rev. A {\bf 72}, 042112 (2005);
I. Peschel and J. Zhao, J. Stat. Mech. P11002 (2005);
R. Orus et al., Phys. Rev. A {\bf 73}, 060303 (2006).}
\bibitem{itzykson} {C. Itzykson and J-M. Drouffe, {\em Statistical field theory vol.2}, Cambridge University Press (1992).}
\bibitem{schollwock}{U. Schollw{\"o}ck and S. R. White, Methods for time dependence in DMRG, in G. G. G. Batrouni and D. Poilblanc (eds.): Effective models for low-dimensional strongly correlated systems, AIP, Melville, New York (2006)}
\bibitem{peps}{F. Verstraete and J. I. Cirac, arxiv.org:cond-mat/0407066 (2004).}
\end{thebibliography}

\eject

{\bf Supplementary information}: Here we outline the calculation of the asymptotic scaling form of the entanglement entropy. The procedure involves writing down the energy density (energy per unit length) of the critical system for states which have only a finite number ($\chi$) of terms in their Schmidt representation. We argue that as long as we are in the scaling limit, all states and hence their corresponding energy density for a fixed $\chi$ depends on only one other parameter for a given system, which can be chosen so that the energy density is minimised. The state thus obtained is the one for which we calculate the entanglement entropy to investigate its scaling with $\chi$.

The entanglement entropy ($S$) in the vicinity of the critical point is given by~\cite{wilczek,vidalent,calabresecardy}.
\begin{equation}
S = \frac{c}{6} \log \xi,
\label{Eq:entdef1}
\end{equation}
where $c$ is the central charge and $\xi$, the correlation length.
The other definition of the entropy is~\cite{calabrese1}
\begin{equation}
S=2b,
\label{Eq:entdef2}
\end{equation}
where $b=-2\log \lambda_{max}$ and $\lambda_{max}^2$ is the largest eigenvalue of the reduced density matrix. $b$ is also the parameter that appears in the Calabrese-Lefevre distribution for the eigenvalues $\lambda^2$ of the reduced density matrix in the vicinity of the critical point~\cite{calabrese1},
\begin{equation}
n(\lambda) = I_0\left(2\sqrt{-b^2-2b\log \lambda} \right).
\label{Eq:CL}
\end{equation}
$I_0(x)$ is the zeroth modified Bessel function. From Eqns.~\ref{Eq:entdef1} and~\ref{Eq:entdef2}, we obtain a relation between $b$ and $\xi$, which is
\begin{equation}
b=\frac{c}{12} \log \xi.
\label{Eq:defb}
\end{equation}

The leading correction to the free energy density ($F$) of a critical point with conformal invariance is (cf. I. Affleck, {\em Phys. Rev. Lett.} {\bf 56}, 746 (1986))
\begin{equation}
F = E_\infty - \frac{\pi c T^2}{6 v},
\label{Eq:FatT}
\end{equation}
where $E_\infty$ is the energy density of the critical state and $v$, the velocity of low energy excitations. Using the thermodynamic identity 
\begin{equation}
E = F - T \left(\frac{\partial F}{\partial T} \right)_L,
\end{equation} 
we can calculate the energy density ($E$) at temperature $T$:
\begin{equation}
E = E_\infty + \frac{\pi c T^2}{6 v}.
\label{Eq:EatT}
\end{equation}
For a one dimensional critical point with conformal invariance, a non-zero temperature $T$ is equivalent to a finite correlation length $\xi$ with $\xi \propto 1/T$. Thus, the energy density ($E$) of a state with a finite $\xi$ at the critical point of a system is given by
\begin{equation}
E = E_\infty + \frac{A}{\xi^2},
\label{Eq:encrit}
\end{equation}
where $A$ is a non-universal constant. The critical state has infinite entanglement entropy and thus requires an infinite number of terms in the Schmidt decomposition to represent. To study the scaling of the entanglement entropy, we allow only a finite number ($\chi$) of terms in the representation of the ground state of the critical system and calculate the value of $S$ as $\chi \rightarrow \infty$. The effect of a finite $\chi$ is to introduce a finite correlation length $\xi$. This implies that the spectrum has a gap $\Delta \sim 1/\xi$. We now argue that another term needs to be added to Eqn.~\ref{Eq:encrit}, for states with a finite $\chi$. To see this we first note that the state $\ket{\psi_0}$ which has the energy density given by Eqn.~\ref{Eq:encrit} for a given $\xi$ has the Schmidt decomposition
\begin{equation}
\ket{\psi_0} = \sum_{n=1}^\infty \lambda_n \ket{\psi^L_n} \ket{\psi^R_n},
\label{Eq:actgrnd}
\end{equation}
where $\ket{\psi^L_n}$ and $\ket{\psi^R_n}$ are states of the two (semi-infinite halves) of the system to the left and right of a given point respectively. $\lambda_n$ has a distribution like in Eqn.~\ref{Eq:CL} with $b$ given in terms of $\xi$ by Eqn.~\ref{Eq:defb} and with the normalization
\begin{equation}
\sum_{n=1}^\infty \lambda_n^2 = 1.
\label{Eq:ketnormal}
\end{equation}
$n$ in Eqns.~\ref{Eq:actgrnd} and~\ref{Eq:ketnormal} goes all the way to infinity as required by Eqn.~\ref{Eq:CL}. However, the state with finite $\chi$ is a truncation of the sum in Eqn.~\ref{Eq:actgrnd} with only $\chi$ terms. This state $\ket{\psi}$ which is an approximation to $\ket{\psi_0}$ is thus
\begin{equation}
\ket{\psi} = \frac{\sum_{n=1}^\chi \lambda_n \ket{\psi^L_n} \ket{\psi^R_n}}{\sqrt{\sum_{n=1}^\chi \lambda_n^2}},
\label{Eq:truegrnd}
\end{equation}
and has an energy density different from the one given by Eqn.~\ref{Eq:encrit}. A knowledge of the energy density $E_0$ of the ground state $\ket{\psi_0}$ and the excitation spectrum, permits us to calculate the energy density $E$ of $\ket{\psi}$ as
\begin{eqnarray}
E &=& E_0 |\bra{\psi_0}\psi \rangle|^2 + E_{ex}\left(1-|\bra{\psi_0}\psi \rangle|^2\right) \cr
&=& E_0 + \left(E_{ex}-E_0\right)\left(1-|\bra{\psi_0}\psi \rangle|^2\right).
\label{Eq:enpert}
\end{eqnarray}
$E_{ex}$ is a measure of the energy density of the excited states and thus $\left(E_{ex}-E_0\right) \sim \Delta \sim 1/\xi$. Using this fact and Eqns.~\ref{Eq:actgrnd} and~\ref{Eq:truegrnd} in Eqn.~\ref{Eq:enpert}, we obtain
\begin{equation}
E = E_0 + \frac{\beta}{\xi}\left(1 - \sum_{n=1}^\chi \lambda_n^2 \right) = E_0 + \frac{\beta}{\xi}P_r(\chi).
\label{Eq:enpert1}
\end{equation}
In the above equation $\beta$ is a non-universal constant and the residual probability $P_r(\chi)$ is given by
\begin{equation}
P_r(\chi) = \sum_{n=\chi+1}^\infty \lambda_n^2.
\label{Eq:resprobdef}
\end{equation}
The $\beta$ term indeed describes an energy cost per length because we have computed the energy cost per truncation, and the number of truncations is linear in length.  We can combine Eqns.~\ref{Eq:encrit} and ~\ref{Eq:enpert1} to obtain the energy per length of the ground state of the critical system for finite $\chi$,
\begin{equation}
E = E_\infty + \frac{A}{\xi^2} + \frac{\beta}{\xi}P_r(\chi).
\label{Eq:totener}
\end{equation}

There is an optimum value of $\xi$ which mimimizes $E$ for a given $\chi$. If $\xi$ is very large, the second term on the right hand side of Eqn.~\ref{Eq:totener} is very small. However, this also means that $b$ is very large from Eqn.~\ref{Eq:defb} and the $\lambda_n$'s fall off very slowly with $n$ from Eqn.~\ref{Eq:CL}. Thus, the residual probability after discarding a finite number of terms is large and so is the third term on the right hand side of Eqn.~\ref{Eq:totener} in comparison to the second term. On the other hand, if $\xi$ is very small, the effect is the opposite with the second term becoming large and the third small. Thus, there is an optimum $\xi$ (and hence $b$) for a given $\chi$  that minimises the energy density. We will determine this value of $b$ in the limit $\chi \rightarrow \infty$. In this limit we can make $n$ a continuous variable and the residual probability
\begin{equation}
P_r(b,\chi) = \int_\chi^\infty \lambda(b,n)^2 dn,
\label{Eq:resprob}
\end{equation}
where we have made explicit the fact that the eigenvalues and hence also the residual probability depend on $b$. We can invert Eqn.~\ref{Eq:CL} to obtain $\lambda (b,n)$. If $\chi$ is large, so is $n$ and we can use the asymptotic form of Eqn.~\ref{Eq:CL}, which is
 \begin{equation}
n(\lambda) = \frac{1}{\sqrt{4\pi\sqrt{-b^2-2b\log \lambda}}}e^{2\sqrt{-b^2-2b\log \lambda}}.
\label{Eq:nasymp}
\end{equation}
Thus
\begin{equation}
\log n(\lambda) = 2\sqrt{-b^2-2b\log \lambda} + \frac{1}{2}\log \left(4\pi\sqrt{-b^2-2b\log \lambda} \right).
\end{equation}
For sufficiently large values of $n$ and hence $\sqrt{-b^2-2b\log \lambda}$, we the second term on the right hand side can be ignored to obtain
\begin{equation}
\lambda(b,n)^2 = e^{-b}e^{-(\log n)^2/4b}.
\label{Eq:lamfromn}
\end{equation}
and
\begin{equation}
P_r(b,\chi)=e^{-b}\int_\chi^\infty e^{-(\log n)^2/4b} dn.
\label{Eq:intPr1}
\end{equation}
From the properties of Gaussian integrals it follows that if $\frac{\log \chi - 2b}{2\sqrt{b}}$ is large, the above integral gives
\begin{equation}
P_r(b,\chi) = \frac{2be^{-b}\chi}{\log \chi - 2b}e^{-(\log \chi)^2/4b}.
\label{Eq:expPr}
\end{equation}
We will see from the final solution that $b \propto \log \chi$ and thus $\frac{\log \chi - 2b}{2\sqrt{b}}$ is indeed large when $\chi$ is large.

We can now determine which state $\ket{\psi}$ minimises the energy $E$ in Eqn.~\ref{Eq:totener}. $\ket{\psi}$ will depend on the $\chi$ values of $\lambda(b,n)$, which in turn depend on the single parameter $b$ as does $\xi$ through Eqn.~\ref{Eq:defb}. Thus, it is sufficient to minimise the energy with respect to $b$. Taking the derivative of the energy in Eqn.~\ref{Eq:totener} and setting it equal to zero gives
\begin{eqnarray}
&\beta \chi e^{-b\left(1+\frac{12}{c}\right)} e^{-(\log \chi)^2/4b}\Big\{\left[\frac{(\log \chi)^2}{4b^2}-1-\frac{12}{c}\right]\frac{2b}{\log \chi - 2b} \cr
&+\frac{2\log \chi}{(\log \chi - 2b)^2}\Big\} - \frac{24A}{c}e^{-\frac{24b}{c}} = 0.
\label{Eq:mincon}
\end{eqnarray}
In the limit of large $\chi$, the above equation has a solution of the type
\begin{equation}
b = \mu \log \chi.
\label{Eq:defmu}
\end{equation}
Substituting $\mu$ from the above equation in Eqn.~\ref{Eq:mincon} lets us calculate its value. The first term in the curly brackets is independent of $\chi$, while the second term is $O(\frac{1}{\log \chi})$ and can be neglected. Thus, the two leading terms are both of the form $\chi$ to an exponent and the two exponents have to be equal for Eqn.~\ref{Eq:mincon} to hold. This gives us the following quadratic equation for $\mu$:
\begin{equation}
(1-2\mu)^2 = \frac{48\mu^2}{c},
\label{Eq:quad}
\end{equation}
which has the solution
\begin{equation}
\mu = \frac{1}{2\left(\sqrt{\frac{12}{c}}+1\right)},
\label{Eq:musol}
\end{equation}
where we have kept the root that is positive for all values of $c$. From Eqns.~\ref{Eq:defb},~\ref{Eq:defmu} and~\ref{Eq:musol}, we can see that
\begin{equation}
\xi = \chi^\kappa,
\label{Eq:xichi}
\end{equation}
where
\begin{equation}
\kappa = \frac{6}{c\left(\sqrt{\frac{12}{c}}+1\right)}.
\label{Eq:defkap}
\end{equation}
Finally, from Eqns.~\ref{Eq:entdef2},~\ref{Eq:defmu} and~\ref{Eq:musol}, we obtain the finite-entanglement scaling of the entropy, which is
\begin{equation}
S = \frac{1}{\left(\sqrt{\frac{12}{c}}+1\right)} \log \chi.
\label{Eq:entscal}
\end{equation}

\end{document}